\documentclass[twocolumn,prb,showpacs]{revtex4}
\AtBeginDocument{
\heavyrulewidth=.08em
\lightrulewidth=.05em
\cmidrulewidth=.03em
\belowrulesep=.65ex
\belowbottomsep=0pt
\aboverulesep=.4ex
\abovetopsep=0pt
\cmidrulesep=\doublerulesep
\cmidrulekern=.5em
\defaultaddspace=.5em
}

\usepackage{amsmath,amssymb}
\usepackage{graphicx}
\usepackage[usenames,dvipsnames]{xcolor}
\usepackage{amsthm}
\usepackage{booktabs}
\usepackage{hyperref}

\definecolor{myred}{HTML}{FF4136}

\newcommand{\etal}{\emph{et al.}}

\hypersetup{colorlinks=true,citecolor=Orange,linkcolor=Red,urlcolor=Black}

\begin{document}

\title{Magnetic excitation spectra of strongly correlated quasi-one dimensional systems: Heisenberg versus Hubbard-like behavior}

\author{A. Nocera}
\affiliation{Computer Science and Mathematics %
Division and Center for Nanophase Materials Sciences, Oak Ridge National Laboratory, %
 \mbox{Oak Ridge, Tennessee 37831}, USA}
 
\author{N. D. Patel}
\affiliation{Department of Physics and Astronomy, The University of Tennessee, Knoxville, 
Tennessee 37996, USA}
\affiliation{Materials Science and Technology Division, Oak Ridge National Laboratory, Oak Ridge, Tennessee 37831, USA}

\author{J. Fernandez-Baca}
\affiliation{Department of Physics and Astronomy, The University of Tennessee, Knoxville, 
Tennessee 37996, USA}
\affiliation{Quantum Condensed Matter Division, Oak Ridge National Laboratory, Oak Ridge, Tennessee 37831, USA}

\author{E. Dagotto}
\affiliation{Department of Physics and Astronomy, The University of Tennessee, Knoxville, 
Tennessee 37996, USA}
\affiliation{Materials Science and Technology Division, Oak Ridge National Laboratory, Oak Ridge, Tennessee 37831, USA}

\author{G. Alvarez}
\affiliation{Computer Science and Mathematics %
Division and Center for Nanophase Materials Sciences, Oak Ridge National Laboratory, %
 \mbox{Oak Ridge, Tennessee 37831}, USA}
\begin{abstract}
We study the effects of charge degrees of freedom on the spin 
excitation dynamics in quasi-one dimensional magnetic materials.
Using the density matrix renormalization group method, we
calculate the dynamical spin structure factor of 
the Hubbard model at half electronic filling
on a chain and on a ladder geometry, and compare the results with those obtained 
using the Heisenberg model, where charge degrees of freedom are considered frozen. 
For both chains and two-leg ladders, we find that the Hubbard model spectrum qualitatively
resembles the Heisenberg spectrum -- with low-energy 
peaks resembling spinonic excitations -- 
already at intermediate on-site repulsion as small as 
$U/t \sim 2-3$, although ratios of peak intensities at different momenta 
continue evolving with increasing $U/t$ converging only 
slowly to the Heisenberg limit.
We discuss the implications of these results for neutron scattering experiments
and we propose criteria to establish the values of $U/t$ of quasi-one dimensional
systems described by one-orbital Hubbard models from experimental information.
\end{abstract}


\maketitle
\newpage

\section{Introduction}
In recent years, we have witnessed a considerable improvement in the 
momentum and frequency resolution
of inelastic neutron scattering techniques, which have been shown to be powerful 
tools to analyze the magnetic excitation dynamics in low dimensional strongly 
correlated materials. A typical example of the accurate agreement 
between theory and experiment is given by the dynamical spin structure factor 
of the one-dimensional spin 1/2 Heisenberg quantum 
magnet $\textrm{KCuF}_{3}$.\cite{re:Lake2013} 

Strongly correlated materials are often described in terms of model Hamiltonians 
where only spin degrees 
of freedom are taken into account---typically the Heisenberg Hamiltonian, which represents 
the main paradigm for quantum magnetism. This is due to the presence of strong electronic 
correlations, which energetically forbid the possibility of double occupation of the outer 
shell orbitals. In these systems, 
charge degrees of freedom are considered \emph{frozen} (or gapped), and 
the low-energy magnetic excitations can be understood in terms of the sole 
spin degrees of freedom.

The idea of using a Heisenberg Hamiltonian as a ``phenomenological'' 
model, even in cases when it is known that it is not fully applicable, 
is not new. For instance, the description of the spin waves in iron 
in the low-energy regime 
has been discussed in terms 
of the Heisenberg model\cite{re:Collins1969,re:Perring1991} since 
the early days of neutron scattering.
Even though this system is clearly itinerant, the Heisenberg model
works
 because it 
captures the essential features of the dispersion relation in this 
low-energy regime. However, it was known that this model 
could not explain the higher energy features 
of the magnetic excitation spectrum, features requiring a more realistic 
treatment that accounts for the itinerant nature of electrons in iron.

One- and quasi-one- dimensional (1D) Mott insulators, 
such as spin chains and ladders, provide an exciting playground for the study of 
strongly correlated quantum states of matter. In 1D, one can observe 
quasi-long-range order states, 
known as Tomonaga-Luttinger liquids,\cite{re:giamarchi2004quantum} 
or phases where correlations between magnetic excitations are short range 
as in the case of Haldane spin-1 chains.\cite{re:Haldane041983}
In ladders, quantum \emph{spin-liquid} phases have properties quite different from those of any 
conventional ferro- or 
antiferromagnet.\cite{re:Dagotto1992} 
In particular, for even number of legs---including the important case of two-leg ladders---the decay of correlations is exponential due to the presence of the spin 
gap.\cite{re:White1994,re:Dagotto1996,re:maekawa1996,re:tranquada2004} 
The existence of this spin gap has been confirmed experimentally  
in the copper oxide SrCu$_{2}$O$_{3}$.\cite{re:Azuma1994} 
Spin ladders have applications ranging from high-temperature
 superconductors\cite{re:dai2012,re:dagotto2013,takahashi2015pressure,re:patel2016} to ultracold atoms.\cite{re:Duan2003,re:Ripoll2004}
Recently, unusual and intriguing physical phenomena were observed in spin systems, 
such as the Bose-Einstein condensation of magnons,\cite{re:zapf2014} the fractionalization of 
spin excitations,\cite{re:Thielemann2009,re:bouillot2011} spinon attraction,\cite{re:Schmidiger2012}
and unusual disorder effects.\cite{re:yu2012}
Excellent agreement between 
theory and experiment has been achieved
\emph{without} considering charge dynamics effects 
in the study of the magnetic 
excitation spectrum, as the example of KCuF$_3$ 
shows.\cite{re:Lake2013} 
Yet 
the effects of the charge degree of freedom cannot always
be neglected, as will be shown in this paper.

An example of the need to consider charge is provided by 
the two-dimensional Mott insulating
cuprate La$_{2}$CuO$_{4}$, where the experimentally observed magnetic dispersion 
departs noticeably from the \emph{pure} Heisenberg form.\cite{re:Coldea2001} 
Starting from the Hubbard model, a perturbation theory in the
electronic hopping term has shown that ring exchange terms appear beyond second order, and they 
are needed to understand the unusual magnetic dispersion and 
thus restore the agreement between theory and experiment. 
This is a direct manifestation that electronic itinerancy or charge dynamics effects are important to understand 
the magnetic excitation spectra of strongly correlated materials.

A recent theoretical study\cite{re:Grage2005} of the dynamical 
spin structure factor in the 1D Hubbard
model~\cite{essler2003} has shown that there are significant charge dynamics effects 
(a transfer of spectral weight) 
due to the coupling of the spin excitations to charge fluctuations 
at low and intermediate values of the Hubbard interaction. 
In this regime, the spin structure factor of the Hubbard model differs from the spectrum 
of the Heisenberg model, that is,
the strong-coupling limit $U/t$ of the half-filled Hubbard model, 
where charge dynamics is suppressed and electrons are completely localized. 
These results have been confirmed in a density matrix 
renormalization group or DMRG\cite{re:White1992,re:White1993} study 
in ref.~\onlinecite{re:Jeckelmann2007}.

Motivated by the mentioned work, this paper studies  
the dynamical spin structure factor of the Heisenberg
and the Hubbard model at half-filling,
 not only for the case of chains, 
but also for two leg ladders.
\emph{The aim of this paper is to provide a quantitative 
criterion to 
determine when a Hubbard model 
description of the material under consideration
 should be preferred to the simpler Heisenberg description.}

The paper is organized as follows. Section~\ref{sec:DSSF} provides an introduction to 
the dynamical spin structure factor and explains how it is computed with DMRG. 
Section~\ref{sec:chains} presents calculations of the dynamical 
spin structure factor of the Heisenberg, 
and of the Hubbard model on a chain. Section~\ref{sec:ladders} extends the comparison 
between the two models studied above in the case of a ladder geometry. 
The last section presents a summary and conclusions.

\section{Dynamical Spin Structure Factor}\label{sec:DSSF}

In this publication we compute spectral 
functions  directly  in  frequency. We follow closely the correction-vector method 
proposed by Kuhner and White in ref.~\onlinecite{re:Kuhner1999}, and calculate  
\begin{equation}\label{eq:Sijw}
S_{i,c}(\omega)=\langle \Psi_{0}|S^{z}_{i} \frac{1}{\omega-H+E_{g}
+i\eta}S^{z}_{c}|\Psi_{0}\rangle,
\end{equation}
at each frequency 
$\omega$ 
for all sites of the lattice, where
 $E_{g}$ is the ground state energy of the time-independent Hamiltonian $H$.
 We consider quasi-one dimensional systems with two possible geometries: 
chains and two-leg ladders.
 In the case of a chain, 
the index $i$ is equal to the only coordinate of the corresponding 
site of the chain. In the case of a two-leg ladder, the index $i\equiv (i_x,i_y)$
corresponds to the two coordinates of the site on the ladder;
$i_{y}=0$ ($i_y=1$) for the lower (upper) leg of the ladder.
The center site $c=L/2-1$ in the case of a chain, and $c\equiv(L/4-1,0)$
in the case of the $L/2 \times 2$ ladder. The chain geometry 
 has $L$ sites, numbered from $0$ to $L-1$; the ladder
 has also $L$ sites in total.
The DMRG correction vector method\cite{re:Kuhner1999} will be used throughout.
Within correction vector we use the Krylov decomposition instead
of conjugate gradient.
A computational advantage\cite{re:nocera2016}
of Krylov decomposition is that the main source of error is given by the Krylov method used for 
the calculation of the correction vector. 
Different frequencies can be computed in parallel, decreasing the 
CPU time needed for the computation of the entire spectrum. The constant $\eta$ has the dimension
of an energy, and constitutes an external parameter for the DMRG 
simulations; $\eta$ controls the broadening of the spectral function peaks.
The above quantity is finally transformed to momentum space as 
\begin{equation}\label{eq:Skw}
S(k,\omega)=\sqrt{\frac{2}{L+1}}\sum_{j=0}^{L-1}\sin((j-c)k)S_{j,c}(\omega+i\eta),
\end{equation}
where the quasi-momenta $k=\frac{\pi n}{L+1}$ with $n=1,..,L$ are appropriate for 
open boundary conditions.
The DMRG implementation used throughout this paper has been 
discussed in ref.~\onlinecite{re:nocera2016}; 
technical details are in the supplemental material, 
which can be found at~\url{https://drive.google.com/open?id=0B4WrP8cGc5JHWXhkcG5wNzk5TDA}. 

\section{One dimensional Chains}\label{sec:chains}

\subsection{Antiferromagnetic Heisenberg model}\label{sec:Heischain}

For a generic quasi-one dimensional geometry, the Hamiltonian of the
antiferromagnetic Heisenberg model is given by
\begin{equation}\label{eq:Heis}
H_{Heis}=\sum_{i,j}J_{i,j}\vec{S}_{i}\cdot\vec{S}_{j},
\end{equation} 
with $\vec{S}=(S^{x},S^{y},S^{z})$. For a chain with open boundaries, 
$J_{i,j} = J$ if $i$ and $j$ are nearest neighbors, and $0$ otherwise.
$J_{i,j}$ will be specialized for ladders in
section~\ref{sec:Heisladder}.

The magnetic excitation spectrum of 
the antiferromagnetic Heisenberg chain 
has been studied thoroughly in the literature, using exact diagonalization,\cite{re:Lin1990}
DMRG\cite{re:Hallberg1995,re:Kuhner1999,re:Jeckelmann2007} and 
analytical approaches.\cite{re:Karbach1997,re:Muller1981}
The ground state energy $E_{GS}=LJ(1/4-\ln2)$, and 
the asymptotic behavior of the static correlation function is\cite{re:Affleck1998}
\begin{equation}\label{eq:HeischainCorr}
\langle GS|\vec{S}_{n}\cdot\vec{S}_{0}|GS\rangle \propto (-1)^{n}\frac{[\ln(n)]^{1/2}}{n},
\end{equation} 
resulting in a weakly diverging static structure factor at $k\simeq\pi$
\begin{equation}\label{eq:HeischainCorrw}
S(k) \propto |\ln|k-\pi||^{\frac{3}{2}}.
\end{equation} 
The Bethe ansatz shows that the manifold of the lowest excited states 
consists of a continuum delimited by a lower and an upper boundary, 
given by the des 
Cloiseaux-Pearson (dCP) dispersions.
The majority of the spectral weight is concentrated in the lower boundary, 
$\omega^{l}(k)=(J\pi/2)\sin(k)$, which is gapless for $q=0$ and $q=\pi$, and repeats itself 
in the other half of the Brillouin zone. Its physical meaning can be explained as follows: 
a flip of a single spin in the antiferromagnetic chain creates a pair of spinons 
(block domain walls in the lattice) having opposite momentum. The spin flip terms in 
the Hamitonian~(\ref{eq:Heis}) move the spinons by two lattice spacings, 
giving to the 
dCP dispersion twice the period. The upper boundary of the excitation manifold is given 
by $\omega^{u}(k)=J\pi|\sin(k/2)|$. The approximate expression 
\begin{equation}
S(k,\omega)=\frac{A}{\sqrt{\omega^2-\omega^{l}(k)^{2}}}\theta(\omega-
\omega^{l}(k))\theta(\omega^{u}(k)-\omega)
\end{equation}
 has been proposed by Muller \etal~in ref.~\onlinecite{re:Muller1981}
 to describe all the features mentioned, where
$A$ is a normalization constant, and $\theta$ is the standard step function.
This ansatz describes very accurately the numerical results obtained 
with correction-vector 
DMRG\cite{re:nocera2016} and time dependent 
DMRG.\cite{re:Lake2013}
A very good agreement between theory (Bethe ansatz) and experiment has been obtained 
using a spin $1/2$ Heisenberg chain model 
description for the compound $\text{CuSO}_{4}\cdot 5 
\text{D}_{2}\text{O}$.\cite{re:Mourigal2013} 
In the supplemental material, which can be found 
at~\url{https://drive.google.com/open?id=0B4WrP8cGc5JHWXhkcG5wNzk5TDA}, 
we have verified that the Krylov method compares well with 
the spectra obtained using a two-spinon exact calculation 
presented in ref.~\onlinecite{re:Karbach1997} for a Heisenberg model. The two spinon solution 
proposed in ref.~\onlinecite{re:Karbach1997} has a similar behavior to the dynamical spectrum  
of the Haldane-Shastry model\cite{re:Haldane1983}, because the latter is in the 
same low-energy universality class of the standard Heisenberg model.

From the Bethe ansatz solution of the Heisenberg model, it follows that 
the $S(k,\omega)$ diverges as
\begin{equation}\label{eq:diverg}
\begin{aligned}
S(k,\omega) &\sim [\omega-\omega^{l}]^{-1/2}\sqrt{\ln[1/(\omega-\omega^{l})]}~\text{for}~k\neq\pi,\\
S(\pi,\omega) &\sim \omega^{-1}\sqrt{\ln(1/\omega)},
\end{aligned}
\end{equation}
as $\omega$ approaches the lower boundary $\omega^{l}(k)$ from above for any $k$ value. 
This divergence has its profound
origin in the Luttinger liquid 
characteristics of the ground state, describes
the instability of the model toward antiferromagnetic ordering, and is expected to 
still be present for 
the Hubbard model on a chain at finite $U$; see next section. For finite size systems, 
one usually cuts off the 
divergences at $\omega-\omega^{l}\simeq 1/L$, so that 
one has peaks of finite height
\begin{equation}\label{eq:max}
\begin{aligned}
\text{max}[S(k,\omega)] &\sim [L\ln(L)]^{1/2}~\text{for}~k\neq\pi,\\
\text{max}[S(\pi,\omega)] &\sim L\ln(L)^{1/2}.
\end{aligned}
\end{equation}
Experimentally, it is not always possible to collect inelastic neutron scattering data 
in the whole relevant range of $k$ and $\omega$, using a single instrument or 
a single configuration. This is because the measurements at low energies 
require higher energy resolution, and because of kinematical constraints of the 
neutron scattering geometry. In these cases much care need to be exercised to 
make direct comparisons between the calculated and measured $S(k,\omega)$ 
over the whole range of $k$ and $\omega$. 

\subsection{Hubbard model and Comparison to Heisenberg}\label{sec:chainHubHeis}

This section reviews and studies
the dynamical spin properties of the Hubbard model with Hamiltonian
\begin{equation}
 \label{eq:H1orb}
H=-\sum\limits_{i,j, \sigma}t_{i,j}c^\dagger_{i,\sigma}c_{j,\sigma}
 + U \sum\limits_{i} n_{i,\uparrow}
n_{i,\downarrow},
\end{equation}
where $U\ge 0$ represents the onsite Coulomb repulsion. 
In the case of a chain with open boundaries, $t_{i,j}=t$ 
if $i$ and $j$ are nearest neighbours, and $0$ otherwise. $t_{i,j}$ 
will be specialized for ladders in section~\ref{sec:Hubladder}.

\begin{figure}
\centering
\includegraphics[width=9.5cm]{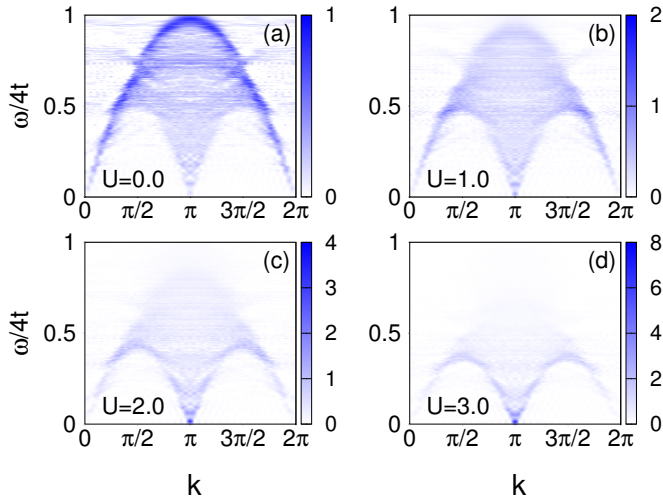}
\caption{(Color online) $S(k,\omega)$ for a Hubbard model 
on a chain of $L=64$ sites at half-filling for different values of $U$, 
as indicated. The number of 
states kept for the DMRG is $m=1000$, $\eta/t=0.05$, while the number of sweeps is $4$.} \label{fig:2}
\end{figure}

Figure~\ref{fig:2} shows the dynamical spin structure factor $S(k,\omega)$ calculated with DMRG 
for a chain of length $L=64$ for different values of the Coulomb repulsion $U$. In this figure, the electronic hopping $t=1$ is assumed as unit of energy, and a broadening of the spectral peaks equal 
to $\eta=0.05$ is used.

\begin{figure}
\centering
\includegraphics[width=8.5cm]{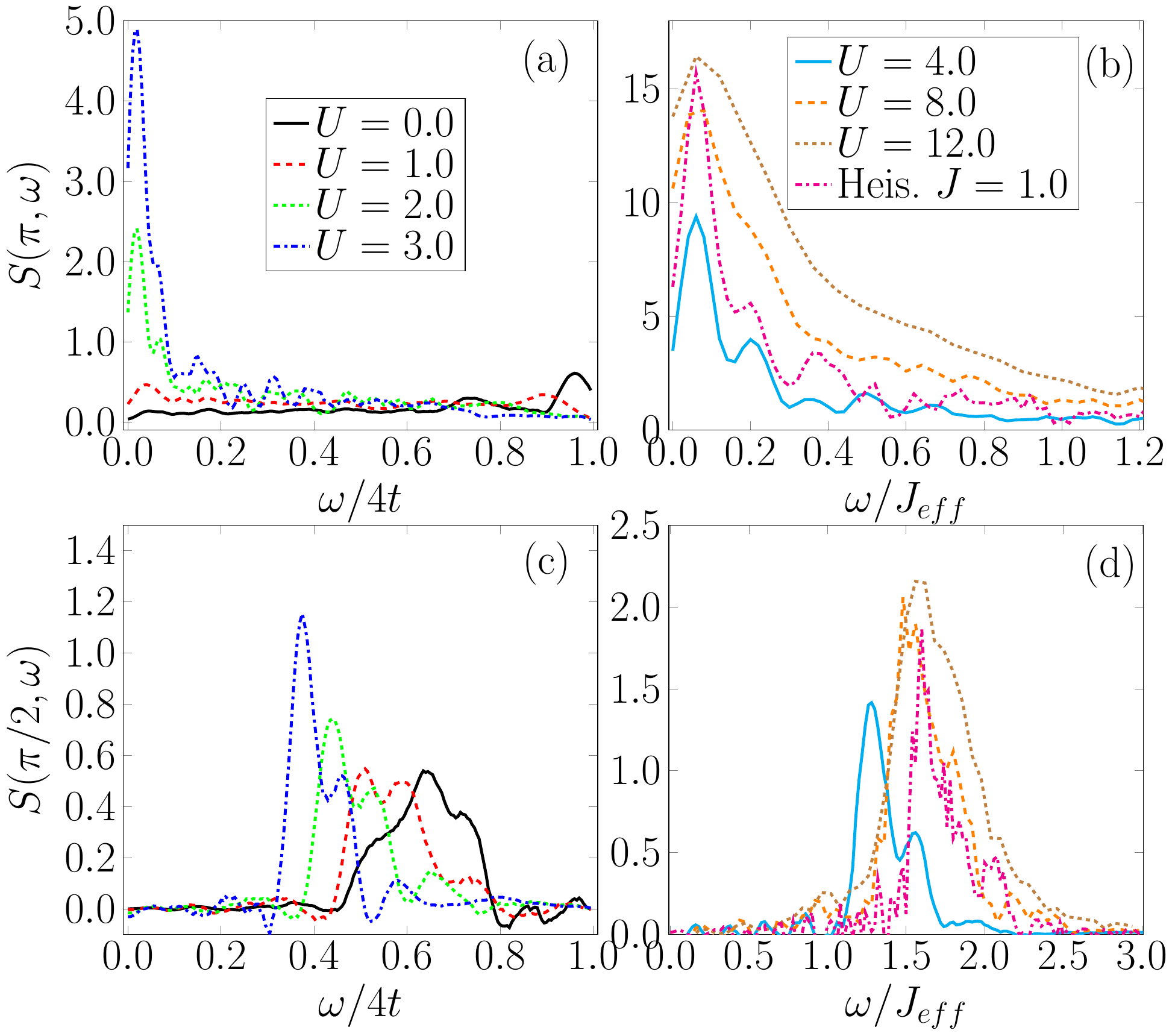}
\caption{(Color online) Panel (a)-(c): cuts of fig.~\ref{fig:2}
as a function of $\omega/(4t)$. Panel (b)-(d): cuts of spectra obtained for 
larger values of $U$,
together with the Heisenberg case cuts. In panels (b,d), on the x-axis we have
plotted frequencies
in units of $\omega/J_{\text{eff}}$, where $J_{\text{eff}}=4t^2/U$. Panels (c) 
and (d) show cuts of the spectrum at $k=\pi/2$.
In panels (a) and (c), we have used $\Delta\omega/t=0.02$ and $\eta/t=0.05$. 
In panels (b) and (d), 
for $U=8$, we have $\Delta\omega/J_{\text{eff}}=0.04$ while for $U=12$ 
$\Delta\omega/J_{\text{eff}}=0.06$. $\eta/t=0.05$
for the Hubbard and $\eta/J=0.05$ for the Heisenberg curves.} \label{fig:3}
\end{figure}

At $U=0$, similarly to the case of the Heisenberg chain analyzed in the previous section, 
the excitation spectrum is enclosed between an upper boundary 
 $\omega_{u}(k)=4t|\sin(k/2)|$ and a lower boundary $\omega_l(k)=2t|\sin(k)|$. The boundaries
 stem from the cosine-like non interacting band structure of the model.
Panels (a-b) and (c-d) of fig.~\ref{fig:3} contain
 cuts at $k=\pi$ and $k=\pi/2$ 
of the spectra shown in fig.~\ref{fig:2}, respectively.
For $k=\pi$, as opposed to the Heisenberg case, the spectral weight is concentrated mostly
at the upper boundary $\omega/(4t)\simeq0.95\simeq\omega_{u}(0)$. Similarly, for $k=\pi/2$ the spectral weight is concentrated in the interval of frequencies $0.5<\omega/(4t)<0.8$ with 
a peak at $\omega/(4t)\simeq0.65\simeq\omega_{u}(\pi/2)$.

When electron-electron interactions
are turned on, the charge dynamics 
manifests itself on the magnetic excitation spectrum as a 
spectral weight redistribution to lower energies.\cite{re:Grage2005}
For $U=1.0$, the dashed (red) curve in fig.~\ref{fig:3}a representing the cut of the spectrum 
at $k=\pi$ shows two weak 
peaks: at $\omega/(4t)\simeq0.9$ and at the lower energy of $\omega/(4t)\simeq0.05$.
A different behavior is observed for the cut at $k=\pi/2$ where the peak position is shifted
to lower energies $\omega/(4t)\simeq0.55$ with an asymmetric triangular shape. 

For $U=2$, the renormalization of the spectral weight has already proceeded 
to lower frequencies 
and the high energy peak characteristic of the non interacting case at $k=\pi$ has almost disappeared:
it is barely visible in our data at $\omega/(4t)\simeq0.8$, as the
short dashed (green) curve in fig.~\ref{fig:3}a shows.
For $k=\pi/2$, the triangular shape spectral feature shows a peak at 
$\omega/(4t)\simeq0.45$.   

 
As noticed in ref.~\onlinecite{re:Jeckelmann2007},
the spectral weight transfer to lower energies is quite fast as a function of $U$, 
and for $U=3$ 
most of the spectral weight is already concentrated very close to the lower boundary of 
the dCP dispersion of the 
Heisenberg model discussed in the previous section; notice the
two very well defined dashed-dotted (blue) line spectral features in panels (a-b) of 
fig.~\ref{fig:3}.
In particular, notice the similarity between the $U=3$ data and the results obtained 
from the Heisenberg model.\cite{re:Lake2013}

For larger values of $U$, shown in panels (b-d) of fig.~\ref{fig:3}, 
the spectral weight redistribution 
continues to approach the Heisenberg-like limit. 
As can be inferred from fig.~\ref{fig:2}, this redistribution
happens for all $k$ values; we have shown them only for
$k=\pi$ and $k=\pi/2$.
The replicas of the main peak at larger frequencies in panel (c) 
of fig.~\ref{fig:3} are artifacts due to finite size effects and 
the use of a tiny broadening $\eta=0.05$.

In these two panels, the cuts are plotted as a function of 
the quantity $\omega/J_{\text{eff}}$.
It is well known that, in the case of half filling and large $U$, 
charge fluctuations are suppressed and the Hubbard model maps onto 
the antiferromagnetic Heisenberg chain with an effective exchange coupling 
constant $J_{\text{eff}}=4t^{2}/U$. 
The results compare very well with those obtained by Benthien and Jeckelmann 
in ref.~\onlinecite{re:Jeckelmann2007}, giving a well defined 
antiferromagnetic peak at $\pi$.

\begin{figure}
\includegraphics[width=8cm]{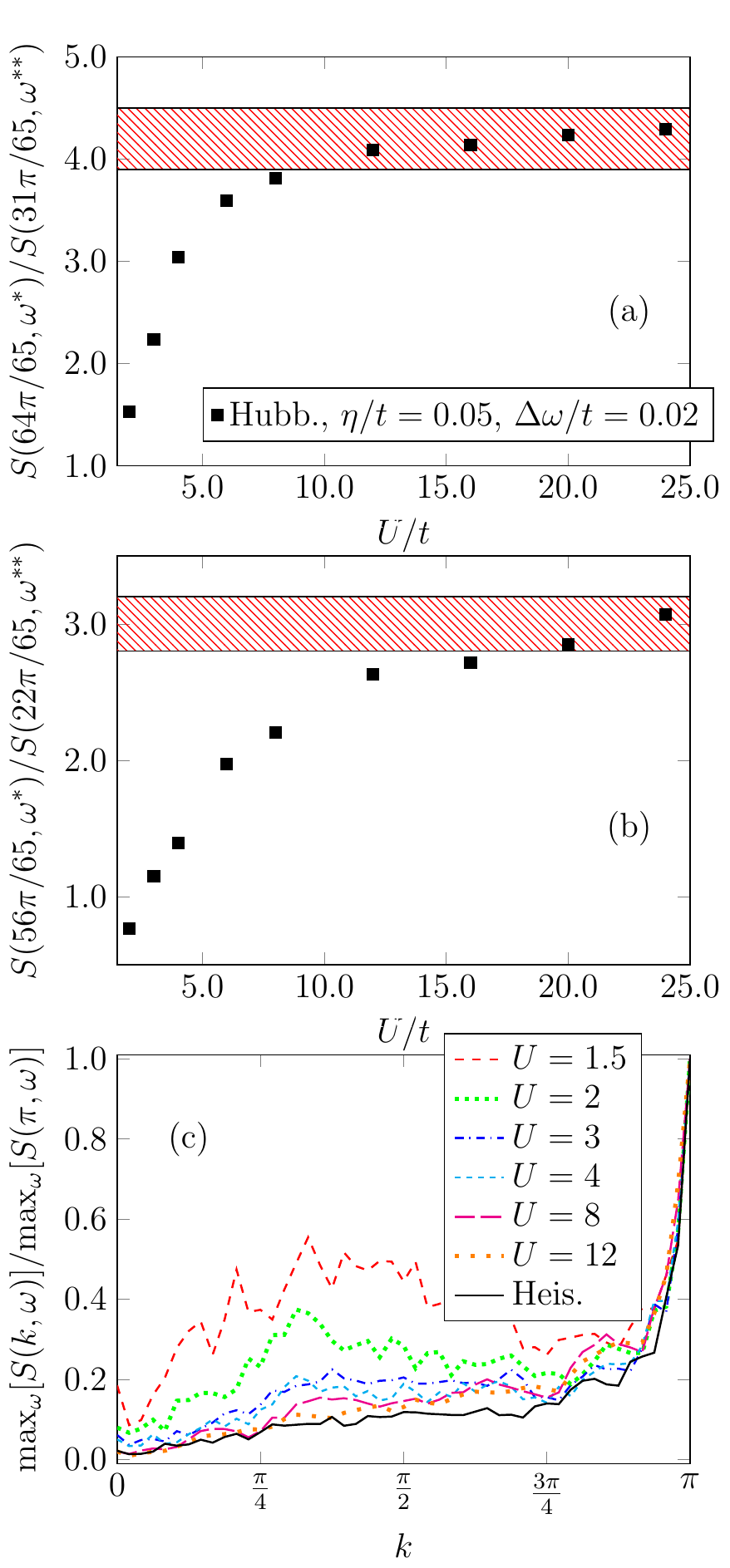}
\caption{(Color online) Panel~(a): Peak to peak ratio 
$S(64\pi/65,\omega^{*})/S(31\pi/65,\omega^{**})$ as a function of 
$U/t$ for the Hubbard model on a chain of $L=64$ sites. 
The ratio obtained for the Heisenberg model is 
indicated by a dashed (red) region, compatible with the error induced by the extrinsic broadening $\eta/J=0.05$ and our resolution in momentum space $k$
(due to finite system size effects). $\omega^{*}$ is the location of 
the peak maximum for $k=64\pi/65$,
$\omega^{**}$ for $k=31\pi/65$.
Panel~(b): Peak to peak ratio 
$S(64\pi/65,\omega^{*})/S(31\pi/65,\omega^{**})$ as a function of 
$U/t$. Here, $\omega^{*}$ is the location of 
the peak maximum for $k=56\pi/65$,
$\omega^{**}$ for $k=22\pi/65$. 
Panel~(c): Peak to peak ratio between the maxima in frequency 
of the spectrum $max_{\omega}[S(k,\omega)]$ and the maximum 
of the spectrum at $k=\pi$ as a function of $k$. 
Hubbard chains of $L=64$ sites and different $U$ 
values are considered, together with the same quantity 
obtained from the Heisenberg model.}\label{fig:3A}
\end{figure}

Panel (a) of fig.~\ref{fig:3A} shows the ratio of the peak maxima for 
$k=64\pi/65$ (closest to $\pi$ for an open chain)\footnote{As also mentioned in the
previous section, the peak at $\pi$ is difficult to observe experimentally,
because the inter-chain coupling is unavoidable in real materials.\cite{re:Lake2005,re:Lake2013}}) and
$k=31\pi/65$ (closest to $\pi/2$ for an open chain) as a function of $U$ for the Hubbard model. 
Panel (b) shows another ratio between the peak maxima for $k=56\pi/65$ (close to $6\pi/7$)
and $k=22\pi/65$ (close to $\pi/3$) as a function of $U$.
In both panels, for $U/t>2$, the ratios
of the peaks maxima of the Hubbard model tend continuously to the
Heisenberg model ratio. The shaded (red) region indicates the interval of 
values for the Heisenberg ratio compatible with our resolution in momentum 
space $\Delta k=2\pi/L$
(due to finite system size effects), the resolution in frequency $\Delta\omega\leq\eta$,
and the extrinsic broadening $\eta$ of the spectral function peaks. 
The width of the shaded region has been estimated as follows. The extrinsic
broadening $\eta$ and the system size of the lattice leads to 
a choice in the mesh in momentum-frequency space. 
Assuming that the peak in the $S(k,\omega)$
for $k=k_0$ has been determined numerically to be at $\omega=\omega_0$,
the half width of the shaded region $\Delta S$ has been estimated as
\begin{equation}\label{eq:DeltaS}
\Delta S = \frac{1}{8}\sum_{m,m'=-1,0,1}|S(k_0+m\Delta k,\omega_0+m'\Delta\omega)-S(k_0,\omega_0)|.
\end{equation}
We have studied chains $L=64$ sites long, 
and for each model have verified that  
the results do not depend significantly on the system size.
The purpose of figure~\ref{fig:3A} is to show that the Hubbard model ratios tend
to the Heisenberg ratio. 
Panel (c) of fig.~\ref{fig:3A} shows the ratio between the spectral 
peak maxima for each value of $k$ in half of the Brilloin zone and the 
same quantity for $k=\pi$. 
The results for $U/(4t)<1$ show a broad peak centered around $k\simeq\pi/3$, a peak not present in the Heisenberg model, and that could thus help distinguish between
the two models in experimental measurements.
This peak gets
suppressed by spectral weight renormalization at larger $U/t$. 

\emph{To conclude this section on chains, a magnetic excitation spectrum qualitatively
 resembling that of the Heisenberg model
is found for the Hubbard model already at $U/t\sim 2$, which is the intermediate
 coupling regime considering
that the bandwidth is $W=4t$. However, fig.~\ref{fig:3A} shows that the ratio
 of intensities between the $k=5\pi/6$ and
$k=\pi/3$ peaks continues evolving with increasing $U/t$ and only slowly
 converges to the Heisenberg limit.
Figure~\ref{fig:3A} thus provides a quantitative criterion 
to decide where a particular one dimensional material
is located in parameter space with regard to the strength 
of its Hubbard coupling.}

\section{Two-leg Ladders}\label{sec:ladders}
This section describes the properties of $S(k,\omega)$ for the same models 
studied in the previous section but in a different geometry---the two leg ladder.

\subsection{Antiferromagnetic Heisenberg ladder}\label{sec:Heisladder}
This section addresses the magnetic excitation spectrum for the spin-$1/2$ Heisenberg 
model on a two-leg ladder. In eq.~(\ref{eq:Heis}), the intra-leg coupling 
$J_{i,j} = J_x$ if $i$ and $j$ 
are nearest neighbors along the $x$ (long) direction; $J_{i,j} = J_y$
if $i$ and $j$ are neighbors along the $y$ (short) direction,
with $J_y$ the inter-leg exchange coupling.
We assume that $J_x,J_y>0$ and in particular $J_x=1$ as our unit of energy. 
This model has attracted much attention in the last twenty years because it represents 
one of the most important paradigms for low dimensional quantum magnetism. 

The Heisenberg model behaves completely different 
on the ladder than on the chain.
As seen in sec.~\ref{sec:Heischain}, the excitation spectrum is gapless at $k\simeq\pi$
for the spin $1/2$ single chain. However, there exists a spin gap in 
the ladder\cite{re:Dagotto1992,re:Barnes1993,re:Rice1993sup,
re:Balents1996,re:Gopalan1994,re:Shelton1996,re:Johnston1996,re:Dagotto1999}---a gap
that has
been experimentally found.\cite{re:Azuma1994,re:Eccleston1994}

Quantum magnetic systems
on ladders have a behavior that can be considered intermediate between 
the one dimensional and two 
dimensional physics.\cite{re:Dagotto1996,re:vuletic2006}
Indeed, both one dimensional and two dimensional magnets have been shown to be gapless. 
It has also been shown that the physics 
of half-integer spins ladders depends on 
the parity of the number of legs.\cite{re:White1994,re:Dagotto1996}

\begin{figure}
\includegraphics[width=9.75cm]{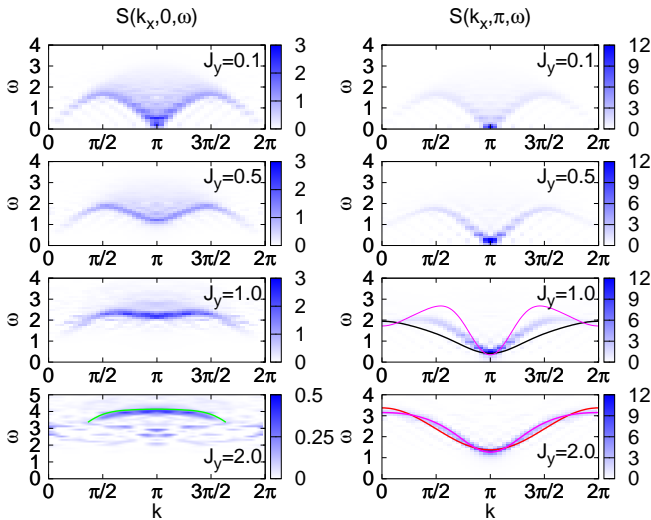}
\caption{(Color online) Left column panels, from top to bottom: 
$S(k_x,0,\omega)$ for the Heisenberg model on a two leg ladder with $L=48\times 2$ sites 
($J_x=1$), for different values of $J_y=0.1, 0.5, 1.0, 2.0$. As before, $m=600$ states 
are kept in the DMRG simulations.
Right column panels: $S(k_x,\pi,\omega)$ for the same set of parameters described above. 
In the panel for $J_y=1$ the dashed line represents the function $\omega_{gap}(k_x)$ 
in eq.~(\ref{eq:Heisladdgap}) obtained in 
ref.~\onlinecite{re:Barnes1993}. The solid lines represent 
the gap dispersion obtained in the plaquette basis in ref.~\onlinecite{re:Piekarewicz1998}. 
In the panel for $J_y=2$ the dashed line represent
eq.~(\ref{eq:Heisladdgap}), 
while the solid line is the gap expression to the 4th order in the 
strong coupling perturbation theory. 
\label{fig:HeisladSqomega}}
\end{figure}

Figure~\ref{fig:HeisladSqomega} shows the dynamical spin structure factor 
for the Heisenberg 
ladder calculated with the DMRG correction 
vector method\cite{re:Kuhner1999} for different values 
of the 
rung exchange coupling $J_y$. 
In the case of the ladder,  
the dynamical structure
factor in momentum space has two components
\begin{equation}\label{eq:Sladw}
\begin{aligned}
S(k_{x},0,\omega)&=\frac{1}{L}\sum_{j=0}^{L/2-1} e^{ik_{x}j}(S(j,0,\omega)+S(j,1,\omega)),\\
S(k_{x},\pi,\omega)&=\frac{1}{L}\sum_{j=0}^{L/2-1} e^{ik_{x}j}(S(j,0,\omega)-S(j,1,\omega)),
\end{aligned}
\end{equation}
because the momentum in the $y$ direction has only two possible 
values: $k_y=0$ or $k_y=\pi$.
 
The dynamical spin structure factor of the Heisenberg ladder has been calculated
both numerically\cite{re:Barnes1993,re:Gopalan1994,re:Yang1998,
re:Sushkov1998, re:Eder1998,re:Suga2002} and
 analytically,\cite{re:Balents1996,re:Sachdev1998} 
 both in the weak coupling limit $J_y\ll1$ and in the strong coupling limit $J_y\gg1$.
 
A simple description of the gap physics can be grasped 
when the strong rung coupling limit $J_y\gg J_x$ is considered. 
Following ref.~\onlinecite{re:Barnes1993}, 
a finite spin gap equal to $J_y$ in the spectrum of spin
excitations obviously appears in the dimer limit, when $J_x=0$.  
Spin excitations can only be produced by promoting a rung singlet to 
a triplet 
at the energy cost $J_y$. These local excitations 
are able to propagate along 
the ladder due to the
perturbation given by the exchange coupling along the legs. 
Using degenerate perturbation theory
in the subspace with one rung promoted to triplet and normal perturbation
 theory on the non-degenerate ground state, one can calculate to 
 order $O(J^{2}_{x}/J_{y})$ the singlet-triplet
dispersion relation
\begin{equation}\label{eq:Heisladdgap}
\omega_{gap}(k_x)=J_{y}\left[1+\frac{J_{x}}{J_y}\cos(k_x)+\frac{3}{4}
\left(\frac{J_x}{J_y}\right)^2\right],
\end{equation}
where $J_{x}$ has been reintroduced for the expression to have consistent energy units,
and to show that the expansion is in $J_x/J_y$.
This gap implies that the spins show no long-range order. 
They usually 
form singlets 
on the rungs and one verifies numerically that the spin correlation decays 
exponentially with distance  
along the legs.\cite{re:Dagotto1992,re:White1994} 

Reference~\onlinecite{re:Piekarewicz1998} carries out the strong coupling expansion 
analytically, including higher order terms, and comparing 
the dimer and plaquette bases. The authors of ref.~\onlinecite{re:Piekarewicz1998} calculate 
the ground state energy and the spin gap to 7th order in the parameter 
$\lambda\equiv J_x/J_y$, and the gap dispersion up to 
4th order. They estimate a radius of convergence $\lambda_c=0.8$ for the pertubative expansion 
by starting from a rung basis, and conclude 
that perturbative expansions in the rung basis are unsuitable for 
dealing with the physically interesting case of isotropic ladders ($\lambda=1$). 
They have explored also the perturbative expansion in a 
$2\times2$ plaquette basis, showing that the radius of convergence 
of the perturbative series can be extended to $\lambda_c=1.25$, 
providing 
a reliable perturbative approach for the isotropic case ($\lambda=1$). 
In the isotropic case, figure~\ref{fig:HeisladSqomega} shows
the gap dispersion obtained 
in ref.~\onlinecite{re:Piekarewicz1998} in the plaquette basis. 
For $J_y=2$, we have included a curve indicating the dispersion obtained in ref.~\onlinecite{re:Piekarewicz1998}
to 4th order of perturbation theory in the rung basis.

Figure~\ref{fig:HeisladSqomega} shows that eq.~(\ref{eq:Heisladdgap})
describes very well the $k_{y}=\pi$ component of 
the dynamical spin structure factor $S(k_{x},\pi,\omega)$.
A strong-coupling calculation reveals that a spin $S=1$
two-particle bound state between two triplets is a possible excitation of 
the system.\cite{re:Sachdev1998} This leads to the presence of a band in 
the spectrum $S(k_{x},0,\omega)$ with a finite range of values of momentum 
transfer $k_{x}$. The dispersion of this band of bound states is almost 
flat, in agreement 
with the analytic expression $\omega_{2triplet}(k)$ of ref.~\onlinecite{re:Monien2001}.

\subsection{Hubbard model and Comparison to Heisenberg}\label{sec:Hubladder}

We now consider a Hubbard model on a two leg ladder geometry. In this case, 
the hopping interaction parameters appearing in 
eq.~(\ref{eq:H1orb}) for the two-leg ladder geometry are as follows: the intra-leg hopping
$t_{i,j}=t_{(i_x,i_{y}),(j_x,i_{y})} = t_x$ is non zero if 
$i_x$ and $j_x$ 
are nearest neighbors on leg $i_y=0$ and $i_y=1$. Furthermore, 
$t_{(i_x,0),(i_x,1)} = t_y$ 
is the rung hopping.
Hubbard ladders are considered as an easier starting point to study the properties of 
the full two dimensional Hubbard model.\cite{re:Dagotto1996}
According to a bosonization approach,\cite{re:Balents1996,re:Fabrizio1992}
the phases of the model can be identified
by  the  number  of  gapless  spin  and  charge  modes,  with the possibility
of having up to two gapless modes in each sector. 
It has been shown that at half-filling the Hubbard model can be found in a 
``C0S0'' phase, where all the charge and spin modes are gapped.
It is however far from obvious that a bosonization picture, which is 
strictly valid when $U\ll t_x$ and at any value of the rung hopping $t_y$,
could completely explain the physics of the problem. Comparison between 
analytical and DMRG calculations has been performed in ref.~\onlinecite{re:White2002}.
Reference~\onlinecite{re:Noack1996} studies
the ground state properties of the model with DMRG, and
reports that phase separation is not found 
in the Hubbard ladder. It also provides evidence 
that the Hubbard model at half-filling, $n=1$, is a spin liquid insulator for 
$t_{y}<2$ at any value of $U$, while it is a band insulator for $t_y>2$. 
A sharp transition between the two phases
turns into a smooth crossover as $U$ is increased.

Unless otherwise stated, in this work we will consider the symmetric $t_x=t_y=1$ case at half 
electronic filling, where the presence of the spin liquid insulator 
phase implies that the spin-spin correlation function decays exponentially 
from the center of the ladder, inducing a gap in the spin sector of the theory.
Away from half-filling, pairing or charge-density wave correlations 
could be dominant. As suggested by the early considerations on Heisenberg and $t-J$ model
ladders, a DMRG study has recently confirmed 
that superconducting correlations become dominant in the limit 
of very small doping.\cite{re:Dolfi2015}
A recent study analyzed the ground state and spectral properties of an 
asymmetric Hubbard ladder,\cite{re:Hohenadler2015} of interest in the context 
of superconducting chains deposited on metallic surfaces.
The dynamical properties of the model have mostly been investigated 
in the half-filled case.\cite{re:Shelton1996,re:Schmidt2005} Away from half-filling, 
 less can be found in the literature.
We should mention a Monte Carlo study\cite{re:Jeckelmann1998} 
and an exact diagonalization study on a t-J ladder.\cite{re:Poilblanc2004} 
Reference~\onlinecite{re:Essler2007} 
uses the connection between the SO(6) Gross-Neveu model
and doped Hubbard ladder to study the spin dynamical structure factor.
Moreover, several low energy analytic descriptions based on bosonization 
and generalized symmetries can be found in the literature.
These studies have focused on the effect of Umklapp processes opening
a gap in the excitation spectrum in the weak 
coupling undoped\cite{re:Liu1998,re:Konik2001,re:Furusaki2002,re:Wu2003} 
and doped cases.\cite{re:Robinson2012}

In the present paper we study the $S(k_x,k_y,\omega)$ of the Hubbard ladder 
with the DMRG using the Krylov approach developed in 
ref.~\onlinecite{re:nocera2016}. 
In the following, we shall consider $t_x=1$ as energy unit. 
\begin{figure}
\centering
\includegraphics[width=9.5cm]{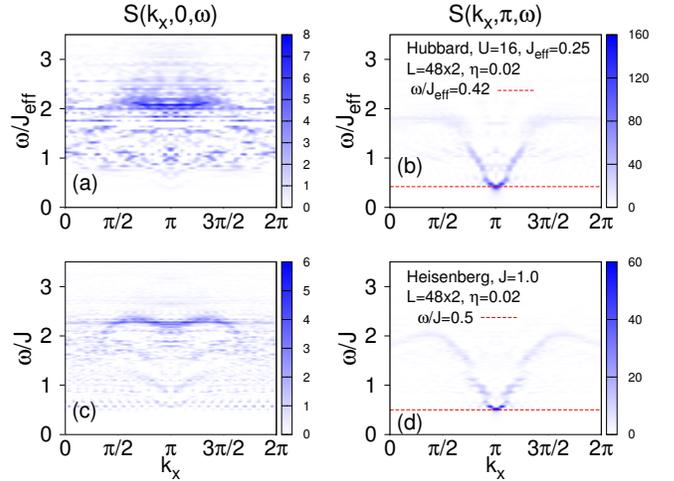}
\caption{(Color online) Left column: $S(k_x,0,\omega)$ component for two leg ladder Hubbard (a)
model with $L=48\times 2$ sites at half-filling, for $t_y=t_x=1$ and $U=16$. 
Panel (c) shows the spectrum for a Heisenberg ladder with $L=48\times 2$ sites and for $J_x=J_y=1$.
In the DMRG simulations, $\eta=\Delta\omega=0.02$ has been considered, 
and up to $m=1000$ DMRG states have been kept in  the numerical simulations. 
Right column: $S(k_x,\pi,\omega)$ component of the 
spectrum for the same parameter values of the left column.} \label{fig:4}
\end{figure} 
\begin{figure}
\centering
\includegraphics[width=9cm]{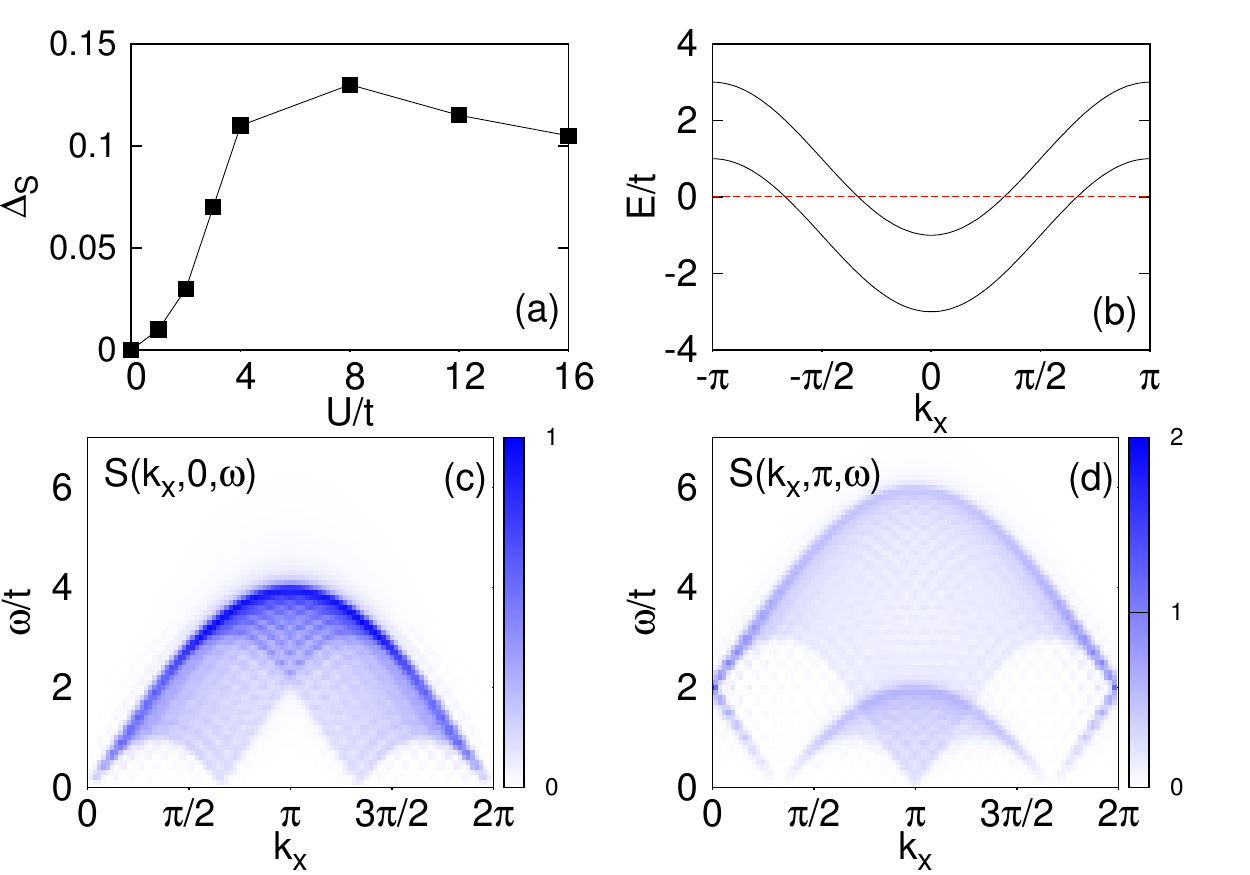}
\caption{(Color online) Panel (a): Spin gap as a function of the on-site Coulomb repulsion $U$ 
extracted from the dynamical spin structure factors for a two leg Hubbard ladder with $L=32\times 2$ 
sites at half-filling. Results agree very well with those obtained in ref.~\onlinecite{re:Noack1996}.
Panel (b): Non-interacting band structure.
Panels (c) and (d): analytical calculation
of the dynamical spin structure factor ($\eta=0.02$) for
a Hubbard ladder with $L=48\times 2$ sites at half-filling, for $t_y=t_x=1$ and $U=0.0$.}\label{fig:5}
\end{figure}
Panels (a) and (b) of fig.~\ref{fig:4} show the two components of the 
dynamical spin structure factor of a
$48\times 2$ ladder with open boundary conditions and $U=16$. 
In the limit of large on-site Coulomb repulsion, the magnetic excitation 
spectrum of the Hubbard model is very similar to that of the 
Heisenberg; this is true for any geometry and is the reason why the spectrum is plotted as a function 
of the quantity $\omega/J_{\text{eff}}$, where the effective exchange coupling constant is $J_{\text{eff}}=4t_y^{2}/U$.

For comparison, panels (c-d) of
the same figure show the $S(k_x,k_y,\omega)$ of an isotropic Heisenberg ladder with $J_x=J_y$
of the same system size.
Both the $k_y=0$ and the $k_y=\pi$ branches of the spectrum of the two models show a strong similarity, 
especially at sufficiently low momenta, 
$|k_x-\pi|<\pi/2$. The $k_y=0$ component of the spectrum is dominated by an almost flat band of
two-triplet bound states, which has robust intensity only for a finite range of momentum 
transfer $k_x$.\cite{re:Sachdev1998,re:Monien2001}
The deviations from a pure Heisenberg model behavior can be appreciable mainly in the spectral weight distribution, which is concentrated at lower momentum transfer in the Hubbard case.
In the $k_y=\pi$ component, the spin gap of the Heisenberg model $\omega_{gap}\simeq0.5$
is well reproduced by the Hubbard result $\omega_{gap}\simeq0.42$.
Even when $U/(4t)=4\gg 1$, deviations from a pure Heisenberg behavior are noticeable
at large momentum transfers, $|k_x|>\pi/2$.

In order to verify the accuracy of our DMRG technique, 
panel (a) of fig.~\ref{fig:5} shows 
the spin gap extracted from the magnetic excitation spectra calculated with DMRG for a 
Hubbard ladder of $L=32\times 2$ sites as a function of $U$; very good agreement 
with the ground stated DMRG results of ref.~\onlinecite{re:Noack1996} is obtained.

Before presenting the results for the interacting Hubbard ladder, it is instructive to analyze 
the properties of the dynamical spin structure factor of a non interacting Hubbard ladder, $U=0$.
The spectrum of the Hamiltonian can be described in terms of bonding ($-$) 
and antibonding ($+$) bands 
\begin{equation}
\epsilon_{a/b}(k_x) = -2t_{x}\cos(k_x)\mp t_{y} ,
\end{equation}
as shown in panel (b) of figure~\ref{fig:5}.
At a given filling $n$ both bands are filled by electrons if the ratio 
$t_{y}/t_{x}$ is less than the critical value $(t_{y}/t_{x})_{c}=1-\cos(\pi n)$. 
At half-filling, $n=1$,
the system is a band insulator for $t_{y}/t_{x}>2$, with a gap equal 
to $2(t_{y}-2t_{x})$,
and a two-band metal otherwise. 
There are four Fermi points: $\pm k_{F_{1}}$ for the bonding, and $\pm k_{F_{2}}$
 for the anti-bonding bands; see panel (b) of fig.~\ref{fig:5}. 
At half-filling $n=1$, one has $k_{F_{1}}+k_{F_{2}}=\pi$ with 
$k_{F_{1}}=\pi/3$ and $k_{F_{2}}=2\pi/3$.

Panels (c) and (d) of fig.~\ref{fig:5} show the $S(k_x,k_y,\omega)$ for a non-interacting isotropic ladder
with $L=48\times2$; $t_y=t_x=1$ is assumed as unit of energy. 
The right panel shows the band structure with bonding and anti-bonding bands.
When the rung hopping is zero, $t_y=0$, then the $0$ and $\pi$ components
of the spin spectrum are equal to each other and $k_{F_{1}}=k_{F_{2}}$, reducing to the case of a 
single chain. When the hopping $t_y$ is increased, the $0$ and $\pi$ components of the 
spin spectrum behave differently, because they encode different excitation processes of the 
spins. Indeed, the processes described by 
the $S(k_x,0,\omega)$ include \emph{intraband} spin excitations, 
analogous to the single chain $S(k,\omega)$, with low energy gapless
contributions at $k_x\simeq0$ and $k_x\simeq2k_{F_{1/2}}$. In contrast, 
the processes described by 
the $S(k_x,\pi,\omega)$ correspond to \emph{interband} transitions between bonding and anti-bonding bands and viceversa,
with characteristic low energy gapless contributions at $k_x\simeq\Delta_{\pm}$, 
where $\Delta_{\pm}=|k_{F_{1}}|\pm |k_{F_{2}}|$. Keeping in mind the above observation, 
panel (c) in fig.~\ref{fig:5} is readily 
understood: low energy spin excitations at $k_x\simeq2k_{F_{1}}=2(\pi-k_{F_{2}})$,
 and $k_x\simeq2k_{F_{2}}=2(\pi-k_{F_{1}})$, together with the usual $k_x\simeq0$ contributions. Except for $t_y=2$, where there are no possible intra-band 
spin excitations, the spectrum upper bound is independent of $t_y$, being $4t_x$. 
The spectral weight of the $S(k_x,0,\omega)$ is concentrated at the upper 
boundary, which follows a sinusoidal behavior equal to that observed in the chain's case.

We now discuss the $\pi$ component of the spectrum, reported in panel (d)
of fig.~\ref{fig:5}. For $t_y>0$, 
 the sinusoidal upper boundary of the spectrum at $t_y=0$ splits into two arcs---arcs 
 separated in frequency by a quantity equal to the bonding-anti-bonding gap, $g=2(t_{y}-2t_{x})$.
Concurrently, characteristic low energy gapless 
contributions at $k_x\simeq\Delta_{\pm}$ appear in the spectrum. As opposed to the $k_y=0$ case, 
the upper bound of the spectrum increases proportionally to $t_y$, because of 
processes exciting spin and charges from the bottom of the bonding band to the top of the anti-bonding band.
As also observed in panel (c) of fig.~\ref{fig:5}, 
the spectrum of the non-interacting case, $U=0$, at $(k_x=\pi,k_y=0)$ 
is gapped up to $\omega_{gap}/(4t)\simeq0.5$, because intra-band excitation with momentum transfer $k_x=\pi$
are not possible due to the band structure topology. For the same reason, electronic 
inter-band excitations are possible instead, and no gap is observed in the $(k_x=\pi,k_y=\pi)$ cut of the spectrum; 
see panel (d) of fig.~\ref{fig:5}.

\begin{figure}
\centering
\includegraphics[width=8.5cm]{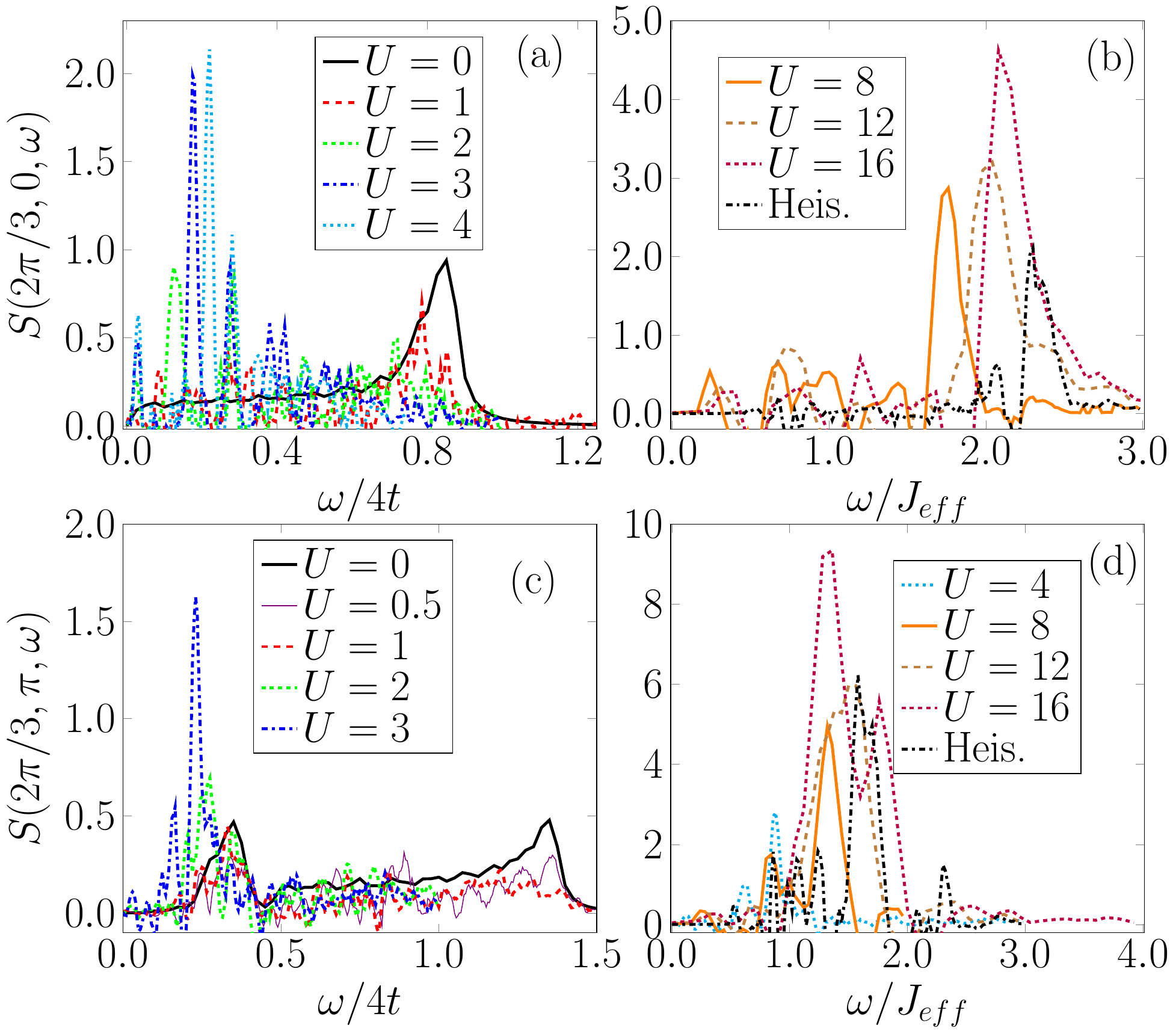}
\caption{(Color online) Cuts of the spin dynamical structure factors 
$S(2\pi/3,0,\omega)$ (panels a and b) and $S(2\pi/3,\pi,\omega)$ (panels c and d) 
for a system with $L=32\times2$ sites 
as a function of $\omega/(4t)$ (panel a and c) and $\omega/J_{\text{eff}}$ (panel b and d), 
at different values of $U$. 
As in the chain case, we have assumed $J_{\text{eff}}=4t^{2}_{y}/U$.
In the DMRG simulations, $\eta=0.02t$ has been considered for the Hubbard model while 
$\eta=0.02J$ for Heisenberg, with $\Delta\omega=0.01$ and with 
up to $m=1000$ states kept in DMRG.}\label{fig:8}
\end{figure}

We are now ready to discuss the results for an interacting Hubbard 
ladder, and compare the results 
to the Heisenberg model. Figure~\ref{fig:8} shows cuts of the $k_y=0$ and
$k_y=\pi$ branches of the magnetic excitation spectrum of a Hubbard ladder 
at $k_x=2\pi/3$ as a function of $U$. The choice of $k_x=2\pi/3$ is motivated
by the gapless behavior observed in the spectrum at $(2\pi/3,0)$ 
and $(2\pi-2\pi/3,0)$ for $U=0$. Figure~\ref{fig:8} shows
a redistribution of spectral weight to lower energy at those $k$ points that 
are gapless in the non-interacting case.
We have already seen this redistribution in the case of the
chain.
The $k_y=0$ branch of the 
spectrum, reported in panel (a) of fig.~\ref{fig:8},
shows exactly this behavior. The peak at $\omega/(4t)\simeq0.9$ is gradually 
suppressed by spectral weight redistribution at lower energy going from $U=0$
to $U=2$. The latter develops a gapped low-energy peak around $\omega/(4t)\simeq0.1$.
The cuts obtained for $U=3$ and $U=4$ have already a pronounced Heisenberg 
like behavior,  with a much suppressed high-energy spectrum and 
the weight concentrated 
around $\omega/(4t)\simeq0.1$. The satellite spectral 
peaks are due to finite size effects and to the choice of a small $\eta=0.02t$. 
The cuts obtained by increasing even further the Hubbard repulsion $U$  
are shown in panel (b) of fig.\ref{fig:8}. Here, one can clearly see
that the results approach a Heisenberg like behavior, characterized by a single
peak at $\omega/J\simeq2.4$. In this panel, as in the case of the chains, 
the cuts are plotted against the ratio $\omega/J_{\text{eff}}$, 
where $J_{eff}=4t_{y}^{2}/U$.

Panels (c) and (d) of figure~\ref{fig:8} show 
the cuts of the $k_y=\pi$ branch of the spectrum 
for different values of $U$ as in the previous panels. 
The redistribution of spectral 
weight to low frequency is also observed here increasing the Coulomb 
repulsion from $U=0$ to $U=1$. Again, the cut of the spectrum for $U=2$ 
presents a suppression of spectral weight at large frequency and 
develops a spectral peak
resembling a Heisenberg-like behavior at $\omega/(4t)\simeq0.25$. The crossover
to a Heisenberg-like behavior is almost complete 
at $U=3$; see the dashed-dotted (blue) curve 
in panel (c). The spectrum approaches the Heisenberg limit for 
very large Coulomb repulsion; see panel (d) of fig.~\ref{fig:8}. As in the chains' case,
the width of the shaded region has been estimated as described in section \ref{sec:chainHubHeis} before eq.~\ref{eq:DeltaS}. 

\begin{figure}
\centering
\includegraphics[width=8cm]{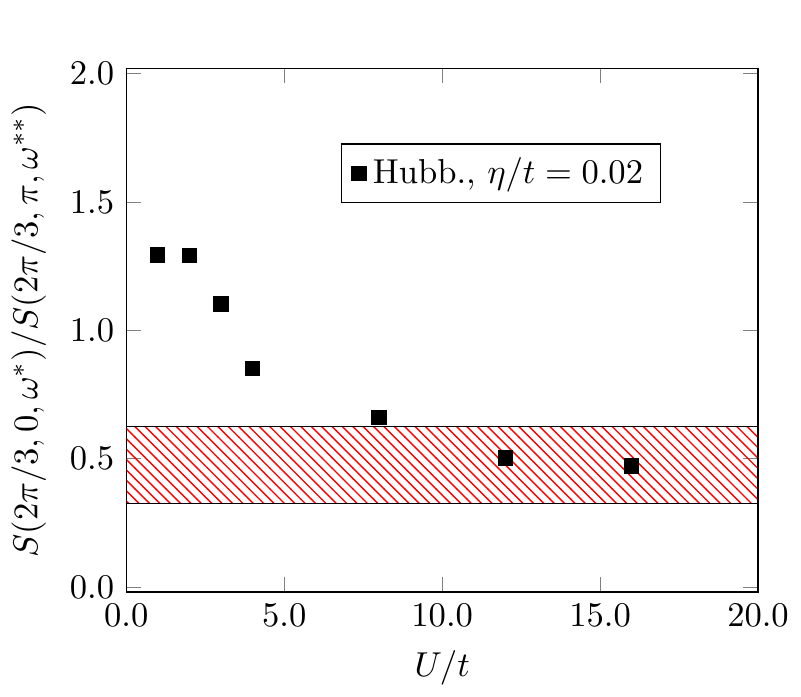}
\caption{(Color online) Peak to peak ratio $S(2\pi/3,0,\omega^{*})/S(2\pi/3,\pi,\omega^{**})$ 
as a function of $U/t$
for the Hubbard model on a ladder of $L=32\times 2$ sites. The ratio obtained 
for the Heisenberg model is 
indicated by a dashed (red) region, compatible with the error induced by 
the extrinsic broadening $\eta/J=0.02$ and our resolution in momentum space $k$
(due to finite system size effects). $\omega^{*}$ is the location of the 
peak maximum for $(k_x,k_y)=(2\pi/3,0)$,
$\omega^{**}$ for $(k_x,k_y)=(2\pi/3,\pi)$. Same parameters' choice as 
in fig.~\ref{fig:8}.}\label{fig:9}
\end{figure}

Figure~\ref{fig:9} shows the ratio
of spectral intensities between the two branches of the spectrum 
at $k_x=2\pi/3$. 
Figure~\ref{fig:9} provides a quantitative criterion 
to decide where a particular quasi-one dimensional material with ladder structure
is located in parameter space with regards to the strength 
of its Hubbard coupling.
\emph{}
Similar to the case of chains studied in the previous section, 
\emph{the peak ratio
evolves continuously with increasing $U/t$, and only slowly
 converges to the Heisenberg limit.
However, qualitatively a Heisenberg-like magnetic excitation spectrum
is found for the Hubbard model already at $U/t\sim2$, which 
for ladders is 
a relatively small coupling regime considering
that the electronic bandwidth is $W=6t$.} 

In the supplemental material,
which can be found at~\url{https://drive.google.com/open?id=0B4WrP8cGc5JHWXhkcG5wNzk5TDA}, we complement 
the study of ladders reported in this section by analyzing 
the cuts of the two branches of the magnetic excitation spectrum 
at the special case $k_x=\pi$, where similar results are obtained.

\section{Summary and Conclusions}\label{sec:conclusions}

In this work, we have compared the dynamical spin structure factor of  
two well known models of strongly correlated materials, the Heisenberg and the 
Hubbard models. By evaluating the 
dynamical spectra we have shown that, both for chains and ladders, 
it is possible to quantitatively identify the range of the on-site 
repulsion strenghts where the Hubbard model resembles that of the Heisenberg model.  Surprisingly, the spectra of the Hubbard model
shows qualitative features that resemble Heisenberg behavior already at relatively
small values of $U$, in particular $U/t \simeq 2-3$ for both chains and ladders. This
explains the success of the Heisenberg model in describing such a wide range of compounds, even 
including metals such
as iron. However, $ratios$ of intensities at various momenta converge slowly to the 
Heisenberg limit and provide an excellent criteria to evaluate the precise value
of $U/t$ from neutron data. 

In fact, current methods and tools of analysis of inelastic neutron scattering data\cite{re:Arnold2014} 
allow for a quantitative evaluation of the magnetic excitation spectrum 
that make possible a direct 
comparison of the relative intensities of the magnetic excitation spectra at 
different wave vectors, as proposed in this paper. This comparison will bring 
considerable light about the applicability of approaches based either on a Heisenberg 
or a Hubbard model. This is in contrast to the earlier days of neutron studies, 
when the information obtained from inelastic neutron scattering was largely limited to 
the peak energy of an excitation, which was plotted against the wave vector in order to 
present the corresponding dispersion relation; see for example ref.~\onlinecite{re:Collins1969}. 
Another important information that is currently available in the neutron inelastic 
scattering spectra is the ``intrinsic'' broadening of the excitations;
this provides essential information about the lifetime of the excitations. The 
intrinsic broadening of the experimentally observed magnetic excitations can be 
evaluated taking into consideration proper corrections for the instrumental resolution.
Future efforts in calculations similar to those in this paper could, in principle, 
account for the broadening mechanisms of the excitations, and provide additional 
information that could be directly compared to neutron scattering experiments.

\section*{Acknowledgments}
This work was conducted at the Center for Nanophase 
Materials Sciences, sponsored by the Scientific User 
Facilities Division (SUFD), BES, DOE, under contract 
with UT-Battelle. A.N. and G.A. acknowledge support by
the Early Career Research program, SUFD, BES, DOE.
N.P. and E.D. were  supported  by  the  National  Science  Foundation  (NSF)
under Grant No. DMR-1404375. N.P. was also  partially  supported  
by  the  U.S.  Department  of  Energy
(DOE),  Office  of  Basic  Energy  Science  (BES),  Materials
Science  and  Engineering  Division. 
Research at ORNL's HFIR and SNS (J. F.-B.) was sponsored 
by the Scientific User Facilities Division, Office of Basic 
Energy Sciences, U.S. Department of Energy.

\clearpage

\bibliography{biblio}

\end{document}